\documentclass[a4paper,12pt]{article}

\pdfoutput=1
\usepackage{jheppub}

\title{Trace Anomaly Matching and Exact Results For Entanglement Entropy}

\author[]{Shamik Banerjee}
\affiliation[]{Kavli Institute for the Physics and Mathematics of the Universe (WPI), Todai Institutes for Advanced Study, The University of Tokyo, Kashiwa, Chiba 277-8583, Japan}

\emailAdd{banerjeeshamik.phy@gmail.com}

\abstract{Following the ideas developed by Komargodski and Schwimmer in \cite{Komargodski:2011vj, Komargodski:2011xv}, we argue that the IR dilaton effective action on the cone computes the entanglement entropy of an even dimensional non-conformal field theory interpolating between a UV and an IR fixed point. We restrict our attention to theories which flow to trivial IR fixed point. We get exact non-perturbative results for the coefficients of the logarithmically divergent term of the entanglement entropy in these field theories in arbitrary even dimensions. The results match precisely with the weak coupling results available in the literature and also with the strong coupling results obtained via holography. We also write the universal terms for field theories which interpolate between two fixed points which are scale invariant but not conformally invariant. }

\begin{document}

\begin{flushright} 
\small{IPMU14-0118} 
\end{flushright}

\maketitle
\flushbottom

\section{Introduction}

 Entanglement entropy is an important and useful quantity which finds applications in many branches of physics. There are very few systems for which one can compute entanglement entropy exactly. Some exact results are known for conformal field theories and for theories which are dual to classical gravity theories in anti-De Sitter space \cite{Srednicki:1993im, Calabrese:2004eu, Calabrese:2009qy, Hartman:2013mia, Faulkner:2013yia, Solodukhin:2008dh, Ryu:2006bv, Ryu:2006ef, Casini:2011kv, Lewkowycz:2013nqa, Faulkner:2013ana, Barrella:2013wja, Cardy:2014jwa, Lewkowycz:2013laa,  Datta:2013hba, Perlmutter:2013gua, Rosenhaus:2014woa}. Holographic method is suited for analyzing strongly coupled field theories and the weakly coupled theories can be accessed via perturbative methods. General non-perturbative techniques for analyzing entanglement entropy in field theories is still lacking. 
 
In recent times new techniques were developed to study renormalization group flow between two fixed points in even dimensions \cite{Komargodski:2011vj, Komargodski:2011xv} \footnote{See also \cite{Luty:2012ww, Schwimmer:2010za, Elvang:2012st}}. The method is based roughly on the fact that in the presence of a background dilaton field, the total scale anomaly of the fluctuating fields and the dilaton matches between the ultraviolet and the infrared. This method is completely non-perturbative and has been used to prove for example the famous $a$-theorem \cite{Komargodski:2011vj, Komargodski:2011xv} in four dimensions. In this note we use this anomaly matching condition to get exact non-perturbative results for entanglement entropy in even dimensional non-conformal field theories. 

\subsection{Results}
One of the richest sources of exact results for entanglement entropy is holography or to be more precise $AdS/CFT$ correspondence. In $AdS/CFT$ one uses the Ryu-Takyanagi conjecture \cite{Ryu:2006bv,Ryu:2006ef, Lewkowycz:2013nqa} to compute the entanglement entropy for strongly coupled large-$N$ gauge theories. On the other hand, entanglement entropy in weakly coupled field theories can be computed by perturbative techniques, although most calculations in the literature are for non-interacting field theories. The only interacting field theories for which entanglement entropy has been studies in great detail are conformal field theories (CFT), in particular two dimensional CFTs. So it will be good to have new tools for calculating entanglement entropy in general interacting field theories not necessarily conformal.  Before we go into the details of the procedure, let us state the results we obtain in this way.

Our results are valid for any even dimensional field theory which is described in the ultraviolet (UV) by a UV-CFT and in the infrared by a trivial IR-CFT. Without any loss of generality we will state the results for four dimensional field theories, but it is straightforward to write down the answer for any even dimension. We have the following setup in mind. We have a conformal field theory in the UV which is deformed by some relevant operator and as a result it flows to some other conformal field theory in the deep IR which is trivial. We denote by $g$ the dimensionless coupling constant of the operator and denote the central charges of a four dimensional conformal field theory by $(a,c)$.

In even dimensions, entanglement entropy has logarithmically divergent pieces. One of our results gives an exact expression for one of the terms. It can be written as,
\begin{equation}
S_E  \supset  - n\frac{\partial}{\partial n}|_{n=1} \ \int_{cone} d^4 x \sqrt h \ (\frac{c_{UV}}{16\pi^2} W^2 - 2 a_{UV} E_4) \ ln\frac{\Lambda_g}{\Lambda}
\end{equation}
The reader should note that this represents a contribution to the entanglement entropy \footnote{Throughout this paper we shall denote entanglement entropy by $S_{E}$.} which arises solely due to the addition of the relevant operator to the UV-CFT. 

Let us now explain the meaning of this formula. The integral is done over a background having conical singularities with the angular excess at the vertices given by $2\pi(n-1)$. The details of the geometry will be discussed later in the paper. $W^2$ and $E_4$ are the Weyl squared term and the Euler density in four dimensions \footnote{Throughout the paper we shall use the conventions of \cite{Myers:2010tj, Myers:2010xs}}. 
$\Lambda$ is the UV-cutoff and $\Lambda_g$ is the renormalization group invariant scale associated with the coupling constant $g$, given by,
\begin{equation}
\Lambda_g = \Lambda e^{- \int\frac{dg}{\beta(g)}}
\end{equation}
$\beta(g)$ is the beta function defined as, $\beta(g) = \Lambda\frac{d}{d\Lambda}g(\Lambda)$. We show later in the paper that this term comes from the universal Weyl anomalous part of the dilaton effective action. 

There are also other logarithmically divergent terms which come from Weyl invariant part of the dilaton effective action. In four dimensions there is only one such term which can be written as, 
\begin{equation}
S_E \supset a_2 \ \Lambda_g^2 \ A_{\Sigma} \ ln\frac{\Lambda_g}{\Lambda}
\end{equation}
where $a_2$ is a coupling constant independent dimensionless number and $A_{\Sigma}$ is the area of the entangling surface. Our method does not allow us to compute the number $a_2$, but it fixes the geometry dependence of this term completely and unambiguously. 

The above two terms are the complete set of logarithmically divergent terms in any four dimensional field theory interpolating between a CFT in the UV and a trivial IR-CFT. Of course we assume that the entangling surface is smooth. These terms give us the correct answer when applied to examples which have been already studied in the literature, see for example \cite{Hertzberg:2010uv, Hertzberg:2012mn, Hung:2011ta, Lewkowycz:2012qr} and references therein.

In higher dimensions one gets more terms. For example in six dimensional theories besides the anomaly and the area term one also gets a term of the form,
\begin{equation}
S_E \supset a_4 \Lambda_g^4 \ ln\frac{\Lambda_g}{\Lambda} \ \int_{\Sigma} R(\Sigma)
\end{equation}
where $R(\Sigma)$ is the Ricci scalar of the entangling surface $\Sigma$ and $a_4$ is a coupling constant independent pure number. We show later in this paper that how one can generate such terms in arbitrary even dimensional field theory. It shows that the lessons obtained for the logarithmically divergent terms for entanglement entropy in $AdS/CFT$ are valid for any arbitrary field theory in even dimensions. 

At the end we study such universal terms in theories which flows form a UV fixed point which is scale invariant but not conformally invariant. To the best of the author's knowledge, this has not been studied in the literature extensively even for the fixed point theories. In this case our result changes slightly. The form of the universal terms depending explicitly on the RG invariant scale $\Lambda_g$ remains unchanged. The dimensionless numbers are expected to change, but in any case our method does not allow us to calculate that. The change in the universal term of entanglement entropy coming from the anomalous part of the dilaton effective action can be calculated exactly. For example, in four dimensions, we get an extra contribution proportional to,
\begin{equation}
n\frac{\partial}{\partial n}|_{n=1} \ e_{UV} \int_{cone} d^4x \sqrt h R^2(h) \  ln\frac{\Lambda_g}{\Lambda}
\end{equation}
where $e_{UV}$ is the coefficient multiplying $R^2$ in the expression for the integrated trace of the energy momentum tensor in a scale but non-conformally invariant theory \cite{Dymarsky:2014zja, Nakayama:2011wq} . So if a theory is described in the UV by a conformally invariant fixed point then then this term will not contribute.

\section{Definition Of Entanglement Entropy}
Let us consider a quantum field theory in $d$ space-time dimensions. We divide the spatial slice into two disjoint parts $A$ and its complement which we call $A^c$.  Let us assume that the system is in the vacuum state denoted by the normalized ket, $|Vac>$. The density matrix which describes the system is, $\rho = |Vac><Vac |$. The Hilbert-space $H$ can be decomposed as the tensor product of the Hilbert spaces, $H_A$ and $H_{A^c}$. The reduced density matrix for the region $A$ is defined as, $\rho_{A} = Tr_{H_{A^{c}}} \rho $. The entanglement entropy for region $A$ is then given by, $S_{A} = -Tr_{H_A} \rho_{A}ln \rho_{A}$. 

The logarithm of the reduced density matrix is difficult to compute directly in a field theory. So one usually uses the replica trick \cite{Holzhey:1994we, Calabrese:2009qy}. Using replica trick one can write the following expression for the entanglement entropy, 
\begin{equation}\label{ent}
S_{A} =  n\frac{\partial}{\partial n} (F(n) - nF(1)) \ |_{n=1}
\end{equation}
where $F(n)$ is the free energy of the Euclidean field theory on a space with conical singularities. The angular excess at each conical singularity is given by $2\pi(n-1)$. We shall talk more about the geometries when we shall need them. 

\section{Entanglement Entropy In Non-conformal Field Theories}
Let us start with the theory of a free scalar field deformed by a mass term. To begin with let us consider two space-time dimensions. The field theory is described by the action,
\begin{equation}\label{s1}
S = \int d^2x \ (\frac{1}{2}(\partial \phi)^{2} + \frac{1}{2} M^2 \phi^2)
\end{equation}
where $M$ is the mass of the scalar field. The action is written in Euclidean signature because that is what is relevant for computing the entanglement entropy. Now if we take the subsystem, for which we are computing the entanglement entropy, to be an infinite half-line then the answer can be shown to be \cite{Calabrese:2004eu,Calabrese:2009qy}, 
\begin{equation}\label{se2}
S_E = -\frac{1}{6} ln \frac{M}{\Lambda} 
\end{equation}
where $\Lambda$ is the ultraviolet cutoff in the field theory.
We can see that the entanglement entropy diverges if we take the ultraviolet cutoff to infinity. The form of the entanglement entropy we have written is of course not unique. In general there are finite terms, which one could add to the answer. These finite additive terms will in general be an infinite series in positive powers of $\frac{M}{\Lambda}$. These terms are non-universal because they can be changed by a finite redefinition of the ultraviolet cutoff $\Lambda$. The universal cutoff independent term is the coefficient of the logarithm, $- \frac{1}{6}$, because it cannot be changed by a finite redefinition of the cutoff. 

Now let us define the massive scalar field theory in the Wilsonian way. We introduce an ultraviolet cutoff $\Lambda_0$ from the beginning and define the coupling constants of the theory at scale $\Lambda_0$. As usual, we define the dimensionless mass $m(\Lambda_0)$ as $m(\Lambda_0)= M\Lambda_0^{-1}$. As we integrate out degrees of freedom, $m(\Lambda_0)$ flows according to the equation,
\begin{equation}\label{se3}
\Lambda \frac{d}{d\Lambda} m = \beta (m) = - m
\end{equation} 
where $\beta(m)$ is the beta function of the coupling $m$ and $\Lambda$ is the floating cutoff. The solution of this equation is of course simple and is given by,
\begin{equation}
m(\Lambda_0)\Lambda_0 = m(\Lambda)\Lambda = M
\end{equation}
The entanglement entropy can be written in terms of the running mass $m(\Lambda)$ as,
\begin{equation}\label{se4}
S_E = - \frac{1}{6} ln \ m(\Lambda)
\end{equation}
The entanglement entropy computed at a different scale $\Lambda'$ will have the universal term written as,
\begin{equation}
S_E' = - \frac{1}{6} ln \ m(\Lambda')
\end{equation}
The difference between the two universal terms is given by,
\begin{equation}
\Delta S_E = - \frac{1}{6} ln \ \frac{m(\Lambda')}{m(\Lambda)} = -\frac{1}{6} ln\frac{\Lambda}{\Lambda'} = \frac{1}{6} \ t
\end{equation}
where we have defined, $\Lambda' = e^t \Lambda$. The important point is that the coefficient of the logarithm is renormalization group invariant. This is a trivial fact in this case but this is useful if the theory is more complicated. In fact this will be the basis of our calculation of the universal terms. 

The example given above was rather trivial. So let us now consider a more general theory in even space-time dimensions equal to $2n$. The theory is defined at scale $\Lambda_0$ and we denote the dimensionless coupling constant by $g(\Lambda_0)$. $g$ couples to some relevant operator of dimension, $\Delta$. Let us further assume that there are some geometric length scales in the problem which we collectively denote by $R$. This length scale may arise either from the background geometry or the geometry of the entangling surface we choose. The entanglement entropy computed in this theory will have logarithmically divergent terms and the coefficient of the logarithm will be the universal term.

Now by dimensional analysis, entanglement entropy will be a function only of the dimensionless coupling $g(\Lambda)$ and $\Lambda R$, where $\Lambda$ is the floating cutoff. So,
\begin{equation}\label{se8}
S_E = S_E (g(\Lambda), \Lambda R)
\end{equation}
The logarithmic term can be written as,
\begin{equation}
S_{univ} = A(g(\Lambda), \Lambda R) \ ln\Lambda
\end{equation}
where $A$ is a finite dimensionless number which can now be a function of various dimensionless quantities characterizing the theory and the background. Now we need to saturate the dimension of $\Lambda$ in the logarithm. In the previous example we did this with the help of the RG invariant mass $M$. So we will follow the same strategy here. We can define the RG invariant mass scale $\Lambda_{g}$ as,
\begin{equation}
\Lambda_g = \Lambda e^{-\int \frac{dg}{\beta(g)}}
\end{equation}
where $\beta(g)$ is the beta function of the coupling $g$. This of course reduces to $M$ in the case of the massive scalar field. So we can write,
\begin{equation}
S_{univ} = - A(g(\Lambda), \Lambda R) \ ln \frac{\Lambda_g}{\Lambda}
\end{equation}
As we have already mentioned, the important point is that the coefficient $A$ should be RG invariant \footnote{In conformal field theories the coefficient of the logarithmic term is Weyl invariant \cite{Solodukhin:2008dh}. The renormalization group invariance of the coefficient of the logarithmically divergent term in a non-conformal theory can be thought of as a generalization of the Weyl invariance in case of a CFT.}. So if we evaluate the entropy at a different scale  $\Lambda'$ then,
\begin{equation}
S_{univ}' = - A(g(\Lambda'), \Lambda' R) \ ln \frac{\Lambda_g'}{\Lambda'}
\end{equation}
Since $A$ and $\Lambda_g$ are RG invariant, we can write, 
\begin{equation}
S_{univ}' = - A(g(\Lambda), \Lambda R) \ ln \frac{\Lambda_g}{\Lambda} + A(g(\Lambda), \Lambda R) t = S_{univ} + A(g(\Lambda), \Lambda R) t
\end{equation}
where we have defined as before, $\Lambda' = e^t\Lambda$. So the logarithmically divergent term is the same at both scales. It easy to see that the RG invariance of $A(g(\Lambda),\Lambda R)$ and $\Lambda_g$ are important for this.

Now since $A$ is RG invariant it has to satisfy the equation,
\begin{equation}
\Lambda \frac{d}{d\Lambda} A(g(\Lambda), \Lambda R) = 0
\end{equation}
We can write the solution of this equation in the form,
\begin{equation}
A(g(\Lambda), \Lambda R) = A(\Lambda_g R)
\end{equation}
This by construction is RG invariant and is the unique solution because the RG equation, $\Lambda \frac{d}{d\Lambda}g = \beta(g)$ has only one conserved quantity associated with it which is $\Lambda_g$. So our strategy for computing the \textit{universal term} will be the following. We will evaluate the entanglement entropy along the RG trajectory. Roughly speaking we will find an expression for $\frac{d}{dt} S_E$ and pick up the RG invariant terms from there. These terms will be the universal terms of the entanglement entropy of the theory. 

Before we leave this section we would like to mention one point. Instead of $\Lambda_g$ we could have saturated the dimension of $\Lambda$ in the logarithm with $R$ which is also RG invariant. This does not make any difference as long as we are interested in the coefficient of the logarithm. Suppose we have a term of the form $B(\Lambda_g R) ln\Lambda R$. We can rewrite this term as,
\begin{equation}
B(\Lambda_g R) \ ln\Lambda R = B(\Lambda_g R) \ ln \frac{\Lambda}{\Lambda_g} + B(\Lambda_g R) \ ln \Lambda_g R 
\end{equation}
Now the second term is not universal and so different ways of writing the logarithm differs by finite non-universal terms. 
Moreover when we compute the derivative $\frac{d}{dt}$ this term drops out because it is manifestly RG invariant. So without any loss of generality we will always write the logarithm as $ln\frac{\Lambda_g}{\Lambda}$.

\section{Two Dimensional Non-conformal Theories}
The two dimensional field theory we will be dealing with is not conformal, but we will assume that in the deep ultraviolet it is described by a $UV-CFT$ and a trivial CFT in the deep IR . For the massive scalar field theory the $UV-CFT$ is just a massless scalar field theory and the $IR-CFT$ is just the trivial gapped theory. Let us also denote by $c_{UV}$ and $c_{IR}$, the ultraviolet and infrared central charges of the $UV-CFT$ and the $IR-CFT$, respectively. In our case $c_{IR} = 0$.  

Suppose we deform the UV-CFT by adding some operator so that it flows. For example, let $O$ be an operator of dimension $\Delta$ and we deform the UV-CFT by this. So, 
\begin{equation}
S = S^{UV}_{CFT} + \int d^2 x \sqrt h \ g(\Lambda_0) \Lambda_0^{2-\Delta} O\label{d1}
\end{equation}
where $h_{ab}$ is a background metric to which the theory is coupled and $g(\Lambda_0)$ is the dimensionless coupling.

Along the RG trajectory the action can be parametrized as, 
\begin{equation}
S(\tau) = S^{UV}_{CFT} + \int d^2 x \sqrt h \ g(\Lambda_0 e^{\tau}) \Lambda_0^{2-\Delta} O\label{d1}
\end{equation}
We want to compute the entanglement entropy for this one parameter family of actions and extract the universal term of the entanglement entropy. What makes life easier is the identification of $\tau$ with a constant background dilaton field.
\subsection{Dilaton And Entanglement Entropy}
In this section we will follow \cite{Komargodski:2011vj, Komargodski:2011xv}. Let us consider renormalization group flow which starts at some ultraviolet(UV) fixed point and end in some infrared(IR) fixed point. Let us assume that the coupling constants have been defined in such a way that $g = 0$ is the UV-CFT. In between the two end-points the theory flows and the conformal symmetry is broken. One can restore the conformal invariance by coupling the theory to a background dilaton field, $\tau(x)$ and making it transform under the conformal transformation simultaneously with the fluctuating fields of the theory. We shall also assume that there is a non-trivial background metric which we will denote by $h_{ab}$. Under a Weyl transformation, $h_{ab} \rightarrow e^{{2\sigma}(x)}h_{ab}$, the dilaton transforms as $\tau \rightarrow \tau + \sigma$. The theory coupled to the background metric and the dilaton is Weyl invariant modulo the conformal anomaly. Now as we integrate out degrees of freedom, we flow to the IR and along the way a dilaton effective action is generated. In the deep IR the theory reaches the IR fixed point and the dilaton decouples from the theory.\footnote{See also \cite{Nakayama:2011wq} for a discussion on this.} One generates a dilaton effective action in the IR in this way. Now the total conformal anomalies in the UV and the IR match in the presence of the background dilaton field. The IR-CFT contributes an anomaly which is not the same as the anomaly of the UV-CFT. So the dilaton effective action must generate an anomaly which will cancel this contribution and give us back the UV-anomaly. So the Weyl-anomalous part of the dilaton effective action is fixed by the anomaly matching and this is the universal part of the dilaton effective action. There is also a part which is diffeomorphism and weyl-invariant and can be written as a functional of the combination, $ \hat h_{ab} = e^{-2\tau}h_{ab}$ which is Weyl invariant.  Now let us see how we can use this observation to calculate entanglement entropy. \footnote{Attempt to prove the a-theorem in four dimensions from entanglement entropy has appeared in the literature \cite{Solodukhin:2013yha}. In this paper we are doing the opposite thing. We are calculating the entanglement entropy using the technology developed for proving the a-theorem in \cite{Komargodski:2011vj, Komargodski:2011xv}. }

\subsection{Entanglement Entropy Of Deformed CFT}
Our deformed theory is not Weyl invariant. This can be made Weyl invariant by coupling to the background dilaton field, $\tau(x)$, as \cite{Komargodski:2011xv},
\begin{equation}
S =  S^{UV}_{CFT} + \int d^2 x \sqrt h \ g(e^{\tau(x)}\Lambda_0) \Lambda_0^{2-\Delta} O \label{dl1}
\end{equation}
Since we are interested in constant rescaling, we can couple to a constant dilaton field $\tau$. Now we can write,
\begin{equation}\label{dl2}
S_{E}(\tau) = n\frac{\partial}{\partial n} (F(n,\tau) - nF(1, \tau)) \ |_{n=1}
\end{equation}
Let us explain the meaning of this formula. $F(n,\tau)$ is the free energy or the effective action computed on the conical space in the presence of the constant background dilaton field coupled to the theory according to Eqn-$4.3$. We want to compute the derivative,
\begin{equation}
\frac{d}{d\tau} S_E(\tau) |_{\tau=0}
\end{equation}
and pick up the RG invariant terms from that. This will give us the universal terms of the entanglement entropy. So 
\begin{equation}\label{main1}
A(\Lambda_g R) = (\frac{d}{d\tau}|_{\tau=0} S_E(\tau))_{RG-Invariant}
\end{equation}
This is the main equation of this paper and we will use this to compute the universal terms.

To compute the derivative we need the dilaton part of the effective action. Now at any intermediate stage of the flow, the dilaton couples nontrivially to the fluctuating fields of the theory. To get rid of this coupling we go to the IR. In the deep IR the theory flows to the IR fixed point and the dilaton decouples. The IR effective action of the dilaton then gives us the complete effective action without coupling to the fluctuating fields of the theory. This of course is valid because by construction free energy is constant along the renormalization group trajectory and the dilaton field is not integrated out because it is a background field.

Now let us say a few words about the term $nF(1,\tau)$. If $F$ is local then the only role of this term is to subtract the bulk contribution from the free energy $F(n)$ and we are left only with the contribution from the tip of the cone. In our case we can replace $F$ with the IR dilaton effective action which is local. So we can neglect the term $nF$, provided we pick up the contribution from the tip of the cone and neglect the contribution from the bulk because that will anyway be  cancelled by the term $nF(1,\tau)$. We will assume this in the rest of the paper.

\section{Another Approach To The Problem}
In this section we will give another argument as to why dilaton is useful for computing the universal terms of the entanglement entropy in field theory. If we are given a conformal field theory in even space-time dimensions then it has conformal anomaly if placed on a curved background geometry. This accounts for the universal terms of the entanglement entropy of the conformal field theory. To be more precise let us assume that the background geometry has a scale associated with it. This scale may come form the geometry of the entangling surface. For example if we take the entangling surface to be a sphere of radius $R$ in flat space then the geometric length scale will be $R$. Then using Eqn-\eqref{ent} we can write \footnote{Please see \cite{Holzhey:1994we, Ryu:2006ef, Solodukhin:2008dh, Myers:2010tj} for more details on this approach.},
\begin{equation}
R\frac{d}{dR} S_E = n\frac{\partial}{\partial n} (R\frac{d}{dR} F(n) - n R\frac{d}{dR}F(1)) |_{n=1}
\end{equation}
Now the variation of the free energy with the scale is given by the integrated trace of the stress tensor of the theory. Here one is supposed to compute the trace on the conical geometry. For a conformal field theory this is given by the integrated conformal anomaly on the cone and one gets the desired universal term in terms of the central charges of the conformal field theory. Since the conformal field theory has no scale other than $R$, in the entanglement entropy this gives rise to a term proportional to $ln \ \Lambda R$ where $\Lambda$ is the ultraviolet cutoff. If the field theory is non-conformal then there are other contributions to the integrated trace of the stress tensor proportional to the beta functions \footnote{The fact that one can compute the universal terms in the deformed theory from the trace of the stress tensor was also mentioned in \cite{Hung:2011ta} }. In order to calculate this contribution let us first couple the non-conformal field theory to a background dilaton field. Since we are only interested in a constant rescaling we can couple it to a constant background dilaton field. Let us now apply the logic of \cite{Komargodski:2011vj, Komargodski:2011xv}. Let $F(h_{ab},\tau)$ denote the free energy of the non-conformal theory coupled to the constant dilaton field $\tau$. If we make an infinitesimal constant Weyl scaling of the metric and also transform the dilaton accordingly then according to \cite{Komargodski:2011vj, Komargodski:2011xv}, the free energy remains unchanged modulo the anomaly of the UV-CFT. So we can write,
\begin{equation}
F(e^{2\epsilon}h_{ab}, \tau + \epsilon) = F(h_{ab},\tau) - \epsilon \int d^{2n}x \sqrt h <T^a_a>_{UV-CFT}
\end{equation}
This equation can be rewritten as,
\begin{equation}
F(e^{2\epsilon}h_{ab}, \tau) = F(h_{ab},\tau - \epsilon) - \epsilon \int d^{2n}x \sqrt h <T^a_a>_{UV-CFT}
\end{equation}
Now the L.H.S can be written as,
\begin{equation}
F(e^{2\epsilon}h_{ab}, \tau) = F(h_{ab},\tau) - \epsilon \int d^{2n}x \sqrt h <T^a_a>_{non-conf + dilaton}
\end{equation}
where $<T^a_a>_{non-conf + dilaton}$ denotes the expectation value of the trace of the stress tensor of the non-conformal theory coupled to the background dilaton field $\tau$. Now comparing the previous two equations we get,
\begin{equation}
\frac{d}{d\tau} F(h_{ab},\tau) |_{\tau=0} = \int d^{2n}x \sqrt h <T^a_a>_{non-conf} - \int d^{2n}x \sqrt h <T^a_a>_{UV-CFT}
\end{equation}
So to compute the trace of the stress tensor in the deformed non-conformal theory we need the dilaton effective action and the complete dilaton effective action is generated by flowing to the deep IR. So we can replace $F$ with the IR dilaton effective action. We can write,
\begin{equation}\label{na1}
\frac{d}{d\tau} F_{IR-dilaton}(h_{ab},\tau) |_{\tau=0} = \int d^{2n}x \sqrt h <T^a_a>_{non-conf} - \int d^{2n}x \sqrt h <T^a_a>_{UV-CFT}
\end{equation}
In this case the IR dilaton effective action has to be calculated on the conical geometry and we need to pick up the RG invariant terms from the R.H.S of Eqn-\ref{na1}. This is precisely the content of Eqn-\ref{main1}. In particular this shows that the dilaton effective action computes the contribution to the entanglement entropy which arises solely due to the addition of the relevant (marginally relevant) term to the UV-CFT. The trace of the energy momentum tensor in the UV-CFT computes the entanglement entropy of the UV-CFT and this contribution is subtracted in the effective action. It will be interesting to see if one could reproduce the results of \cite{Liu:2012eea} using this approach.

\section{Calculations In Two Dimensional Theories}
\subsection{Conformal Field Theories On Cones}
We have seen that in order to compute the entanglement entropy we need to know the infrared dilaton effective action for a constant dilaton field on a cone. A constant dilaton field couples to the integrated trace of the energy-momentum tensor. So due to anomaly matching, the Weyl non invariant universal part of the IR dilaton effective action can be written down once we know the integrated trace of the energy-momentum tensor of a conformal field theory on the cone. For a two dimensional conformal field theory with central charge $c$, the integrated trace of the energy momentum tensor on a flat cone with angular excess $2\pi(n-1)$, can be written as \cite{Holzhey:1994we, Calabrese:2004eu, Calabrese:2009qy}, 
\begin{equation}\label{cone1}
\int_{cone} \sqrt h <T^{\mu}_{\mu}> = \frac{c(n)}{24\pi} \int_{cone} \sqrt h R(h)
\end{equation}
where 
\begin{equation}\label{cone2}
c(n) = \frac{c}{2} (1+ \frac{1}{n})
\end{equation}
and $R(h)$ is the Ricci scalar of the cone. Since we have assumed a flat cone, there is no bulk contribution to the trace and all the contribution arises from the tip of the conical singularities.

Now we would like to make a few comments on this formula. This formula shows that at least in two dimensions, the contribution of the conical singularity to the integrated trace can be written in the standard way except that the central charge $c$ is replaced by an effective central charge $c(n)$, which is a non-trivial function of $n$. As $n\rightarrow 1$, $c(n)\rightarrow c$. In fact we can expand $c(n)$ around $n=1$ and write,
\begin{equation}\label{cone3}
c(n) = c + c_1 (n-1) + ......
\end{equation}
Now suppose we have a flat cone with only one conical singularity. Then \cite{Fursaev:1995ef},
\begin{equation}\label{cone4}
\int_{cone} \sqrt h R(h) = 4\pi(1-n)
\end{equation}
So we can write,
\begin{equation}\label{cone5}
\int_{cone} \sqrt h <T^{\mu}_{\mu}> = \frac{c}{24\pi} \int_{cone} \sqrt h R(h) + O((n-1)^2)
\end{equation}
In order to compute the entanglement entropy we need to differentiate with respect to $n$ and set $n=1$. So the terms of $O((n-1)^2)$ do not contribute and for the purpose of calculating entanglement entropy we can simply replace $c(n)$ with $c$. In fact we shall assume that the following holds for higher dimensional conformal field theories as well. The contribution of the conical singularities to the integrated trace of the energy-momentum tensor is given by the same anomaly polynomial as in the case of a non-singular curved manifold, except that the central charges are replaced by their effective values. The effective central charges are analytic at $n=1$ and so for the purpose of calculating entanglement entropy we can replace them with their values at $n=1$. This is an assumption which should be proved rigorously. The validity of this assumption has been checked in many ways in holographic computations of entanglement entropy in conformal field theories and perturbative calculations in weakly coupled field theories. 

\subsection{Semi-Infinite Line In Two Dimensions}
In two dimensions the universal (Weyl non-invariant) part  of the dilaton effective action in the IR for a constant dilaton filed is given by \footnote{To write down the exact dilaton effective action for arbitrary $n$ we have to replace $c$ with $c(n)$. As we have already mentioned, for the purpose of computing entanglement entropy we can as well work with the standard central charges.},
\begin{equation}\label{ea1}
F_{IR}(n,\tau) = -\frac{c_{UV}}{24\pi} \ \tau  \int_{cone} \sqrt h R(h)
\end{equation}
For the infinite half-line in two dimensions the geometry is that of a flat cone with only one conical singularity. Now using Eqn-\ref{dl2} and Eqn-\ref{cone4} we get,
\begin{equation}\label{ea2}
\frac{d}{d\tau} S_E |_{\tau=0} = \frac{c_{UV}}{6} 
\end{equation}
For the massive scalar field the answer reduces to the answer we have already stated,
\begin{equation}\label{ea4}
\frac{d}{d\tau} S_E(\tau)|_{\tau=0} = \frac{1}{6}
\end{equation}
where we have used the fact that for a massive scalar, $c_{UV} = 1$ and $c_{IR} =0$. These universal terms match precisely with the known result \cite{Calabrese:2004eu, Calabrese:2009qy} in two dimensions. 

We can also write it in the more conventional form as,
\begin{equation}\label{ea8}
S_E = - \frac{c_{UV}}{6} \  ln \frac{\Lambda_g}{\Lambda}
\end{equation}
 Here we can see that the mass has been replaced by the renormalization group invariant scale $\Lambda_g$. Of course in the case of a massive scalar field of mass $M$, $\Lambda_g = M$.
\subsection{Contribution From The Weyl Invariant Part of the Dilaton Effective Action}
The dilaton effective action in the infrared also has a part which is Weyl and diffeomorphism invariant. As we have already mentioned, this part of the effective action can be written as a functional of the combination, $\hat h_{ab} = e^{-2\tau} h_{ab}$. Since the dilaton effective action is local, this can be written in terms of integrals of local densities built out of the tensor $\hat h_{ab}$. Now the infinite half-line has no geometric scale associated to it and so the only term which can contribute to the universal term is given by,
\begin{equation}
\int_{cone} d^2x \sqrt {\hat h} R(\hat h) 
\end{equation} 
Since this term is topological in two dimensions this does not couple to the dilaton. So we do not get any new contribution from the Weyl invariant part of the dilaton effective action. 

\section{New Results For Higher Dimensional Non-conformal Field Theories}

\subsection{Four Dimensions}
Let us first calculate the terms captured by the universal anomalous part of the dilaton effective action. 
The universal part of the effective action for a constant dilaton field in four dimensions can be written in terms of the two central charges, $a$ and $c$, as \footnote{We will follow the conventions of \cite{Myers:2010tj,Myers:2010xs}.}, 
\begin{equation}
F(n,\tau) = - \tau \int_{cone} d^4 x \sqrt h \ (\frac{c_{UV}}{16\pi^2} W^2 - 2 a_{UV} E_4)
\end{equation}
where $W^2$ and $E_4$ are the Weyl tensor squared and the four dimensional Euler density defined as,
\begin{equation}
W^2 = R_{abcd}R^{abcd} - 2R_{ab}R^{ab} + \frac{1}{3} R^2
\end{equation}
\begin{equation}
E_4 = \frac{1}{32\pi^2} (R_{abcd}R^{abcd} - 4R_{ab}R^{ab} + R^2)
\end{equation}
Since we are interested in entanglement entropy, the metric $h_{ab}$ can be taken as the metric of a regularized conical space with infinitesimal angular excess at the singular points. This space depends on our choice of the spatial geometry and the entangling surface. Let us work it out in some simple examples. 
\subsection{Spherical Wave-Guide Geometry} In this case the spatial geometry is that of a cylinder $S^2 \times R^1$ and the entangling surface is just a sphere \cite{Lewkowycz:2012qr}. The full Euclidean geometry is $R^2 \times S^2$ where the $R^1$ comes from time. The geometry of the conical space is just the direct product of a two dimensional flat cone and the sphere $S^2$. In this case the integral can be done easily \cite{Fursaev:1995ef} and we finally get,
\begin{equation}
\frac{d}{d\tau} S_E(\tau)|_{\tau=0} \supset -4 (a_{UV} - \frac{c_{UV}}{3})
\end{equation}
Note that this result is completely non-perturbative and holds for any field theory which interpolates between a $UV-CFT$ and a trivial $IR-CFT$. In more conventional terms,
\begin{equation}
S_E \supset 4(a_{UV} - \frac{c_{UV}}{3}) \ ln\frac{\Lambda_g}{\Lambda}
\end{equation}
This matches precisely with the answer of \cite{Lewkowycz:2012qr}. We have not written an equality sign because there are other universal terms which depend on the scale of the background geometry. 


\subsection{Spherical Entangling Surface in Flat Space}
Let us take a spherical entangling surface of radius $R$ in flat space. In this case the geometry is that of a squashed cone \cite{Fursaev:2013fta} and the relevant integrals can be calculated easily. The final answer in this case is,
\begin{equation}
\frac{d}{d\tau} S_E(\tau)|_{\tau=0} \supset -4 a_{UV}
\end{equation}
So,
\begin{equation}
S_E \supset 4 a_{UV} \ ln\frac{\Lambda_g}{\Lambda}
\end{equation}
\subsection{Cylindrical Entangling Surface in Flat Space}
We take the entangling surface as a cylinder of radius $R$ and length $L$. We take $L$ to be large but finite in order to regulate the entropy. In this case the answer turns out to be, 
\begin{equation}
 \frac{d}{d\tau} S_E(\tau)|_{\tau=0} \supset - \frac{L}{2R} \ c_{UV}
\end{equation}
So,
\begin{equation}
S_E \supset \frac{L}{2R} c_{UV} \ ln \frac{\Lambda_g}{\Lambda}
\end{equation}





\section{Universal Terms Of The Second Kind Or Less Universal Terms}
Universal terms of the second kind arise from the Weyl-invariant part of the IR dilaton effective action. Since the weyl invariant part of the IR dilaton effective action can depend on the details of the renormalization group flow, the universal terms arising from that is less universal in some sense. No simple argument like anomaly matching can give us the coefficients of the universal terms of the second kind. In this case the usefulness of this approach is that it gives us a new organizing principle. First of all since the dilaton effective action is local, the universal terms of the second kind can all be expressed as the integral of a local density on the entangling surface. Although this can be checked in perturbative calculations in weakly coupled field theories, this is not a trivial statement for a strongly coupled field theory. In fact this was observed in holographic computations of entanglement entropy \cite{Hung:2011ta}. So the locality of the dilaton effective action explains this observation. 

We have seen that the universal coefficient of the logarithmically divergent term is RG invariant and can be written as a function of the combination $\Lambda_g R$, where $\Lambda_g$ is the RG invariant scale. So far we have seen only the $R$ independent part of the universal term, for example $c_{UV}$. Now to see the terms which genuinely depend on the dimensionful parameter $\Lambda_g$ we need to look at the Weyl invariant part of the dilaton effective action. Let us now do that. The following analysis of the universal terms is particularly for four dimensions, but the reader will notice that this can be easily generalized to any even dimensions.

\subsection{Weyl Invariant Dilaton Action}
We have already mentioned that the Weyl invariant part of the dilaton effective action can be written as a functional of the Weyl invariant tensor, $ \hat h_{ab} = e^{-2\tau}h_{ab}$. In our case the background dilaton field $\tau(x)$ is a constant. The dilaton effective action can be expanded in terms of tensors built out of $\hat h$. Let us arrange these terms in order of increasing mass dimensions of the integrand. 

The first term is 
\begin{equation}
\int_{cone} d^4x \sqrt {\hat h} = e^{-4\tau} \int_{cone} d^4x \sqrt h
\end{equation}
This term does not contribute to the entanglement entropy because the volume of the cone does not get any contribution from the tip.

The second term is,
\begin{equation}
\int_{cone} d^4x \sqrt {\hat h} R(\hat h) = e^{-2\tau} \int_{cone} d^4x \sqrt h R(h)
\end{equation}
This term will generically contribute. For example let us take the spherical wave-guide geometry. In this geometry the integral evaluates to, $4\pi(1-n)A_{S^2}$ where $A_{S^2}$ is the area of the sphere \cite{Fursaev:1995ef}. We can replace the sphere with any two dimensional manifold and we will get the area of that manifold replacing $A_{S^2}$. Since this term has mass dimension $-2$ we need some mass parameter to saturate the dimension. If this term appears as a universal coefficient of the logarithmically divergent term then according to our previous argument this mass scale can only be the RG invariant scale $\Lambda_g$. 
So we get a contribution to the entanglement entropy of the form,
\begin{equation}
S_E \supset a_2 \Lambda_g^2 A_{\Sigma} \ ln\frac{\Lambda_g}{\Lambda}
\end{equation}
where $\Sigma$ is the two dimensional entangling surface. It should be noted that there is no large or small $\Lambda_g$ expansion involved here. The dependence is fixed completely by the locality of the dilaton effective action and dimensional analysis. Since $a_2$ is dimensionless, it is a pure number and a universal number characterizing the field theory. This type of terms arise in weak coupling for a massive scalar and fermion field \cite{Hertzberg:2010uv,Lewkowycz:2012qr}  and in holographic computations with strongly coupled field theories \cite{Hung:2011ta}. In general it is difficult to determine the precise geometrical term which gives rise to this contribution. We can see the usefulness of this method in determining the structure of these terms. The present method also makes it clear that these terms arise not only for weakly coupled filed theory or strongly coupled field theory, but for any field theory. 

The dimension four terms can be written as linear combinations of $R^2(\hat h)$, $R^2_{ab}(\hat h)$ and $R^2_{abcd}(\hat h)$. So a general dimension four term in the dilaton effective action has the structure,
\begin{equation}
\int_{cone} d^4x \sqrt{\hat h} (A R^2(\hat h) + B R^2_{ab}(\hat h) +C R^2_{abcd}(\hat h))
\end{equation}
where $A$, $B$ and $C$ are dimensionless constants. It is easy to see that this term does not couple to a constant dilaton and so does not contribute to the universal term. In fact this is the reason why the universal term of the first kind does not get any contribution from the Weyl invariant part of the dilaton effective action. This term is marginal and if this term contributed to the universal term then it would have changed our answer for the universal term of the first kind. 

In higher dimensions things go exactly in the same way and we will not discuss this further. 

\section{An Example: Massive Scalar Field On The Spherical Wave-Guide Geometry }

To see how this method works let us analyze the example of the spherical wave-guide for which the entanglement entropy was calculated in \cite{Lewkowycz:2012qr} by using the heat-kernel technique. So the geometry of the Euclideanized space is just $R^2 \times S^{N-2}$ and the spatial slice has the geometry of a cylinder, $R^1\times S^{N-2}$. We take the radius of the sphere to be $R$. The entanglement entropy is computed for the half-space $R^{+}\times S^{N-2}$. For the computation of the entanglement entropy we need to work on the conical space $C^2_n\times S^{N-2}$, where $C^2_n$ is the two dimensional flat cone with only one singularity with angular excess $2\pi(n-1)$. As we have alraedy described we need to consider $n$ only in an infinitesimal neighborhood of $1$.

One of the examples \cite{Lewkowycz:2012qr} considered is a massive scalar field with non-minimal coupling to the curvature of the background geometry, described by the action
\begin{equation}
S = \frac{1}{2} \int_{R^2\times S^{N-2}} d^{N}x \sqrt h \ ((\nabla\phi)^2 + M^2 \phi^2 + \xi R(h) \phi^2)
\end{equation}
where $R(h)$ is the Ricci scalar of the background metric $h_{ab}$. The coefficient $\xi$ is the curvature coupling which for a conformally  coupled scalar is given by,
\begin{equation}
\xi =  \frac{N-2}{4(N-1)}
\end{equation}
Now let us define,
\begin{equation}
\xi' = \xi - \frac{N-2}{4(N-1)}
\end{equation}
So $\xi' = 0$ for the conformally coupled scalar. Now we can write the action in the form,
\begin{equation}
S = S_{UV-CFT} + \frac{1}{2} \int_{R^2\times S^{N-2}} d^{N}x \sqrt h \ \tilde M^2 \phi^2
\end{equation}
where,
\begin{equation}
S_{UV-CFT} =  \frac{1}{2} \int_{R^2\times S^{N-2}} d^{N}x \sqrt h \ ((\nabla\phi)^2 + \frac{N-2}{4(N-1)} R(h) \phi^2)
\end{equation}
and
\begin{equation}
\tilde M^2 = M^2 + \xi' R(h)
\end{equation}
For this particular geometry,
\begin{equation}
R(h) = \frac{(N-2)(N-3)}{R^2} = R(S^{N-2})
\end{equation}
The renormalization group invariant scale in this case is given by,
\begin{equation}
\Lambda_{\tilde M} = \tilde M
\end{equation}

\subsection{N=4}
In case of four dimensional space, according to our discussion in the previous section there is only one term of the form,
\begin{equation}\label{sw0}
S_E \supset a_2 \tilde M^2 \ ln\frac{\Lambda_{\tilde M}}{\Lambda} \int_{cone} R 
\end{equation}
So we get a contribution of the form,
\begin{equation}\label{sw1}
(M^2 + \xi' R(S^2)) A_{S^2}
\end{equation}
where $\xi' = \xi - \frac{1}{6}$. The reader should note that there is no large or small mass expansion involved in this argument. This approach does not allow us to compute the precise value of the coefficient $a_2$, because the anomaly matching condition does not fix the coefficients of the Weyl invariant part of the dilaton effective action, but we can see that it completely determines the geometrical structure of the term. We do not have terms cubic or higher orders in the curvature because those will have to be multiplied by negative powers of the renormalization group invariant scale. This cannot happen because in that case the dilaton effective action will diverge in the conformal limit and we do not expect it to do so. So higher curvature terms are ruled out. 

The form of our result matches precisely with the result of \cite{Lewkowycz:2012qr} obtained by heat-kernel technique, modulo the coefficient $a_2$. In spite of that we would like to interpret this result in a slightly different way. \cite{Lewkowycz:2012qr} wrote the term $\xi' R(S^2)A_{S^2}$ as,
\begin{equation}\label{sw2}
\xi' R(S^2) A_{S^2} = \xi' \int_{S^2} R(S^2)
\end{equation}
and interpreted it as a curvature contribution in four dimensions. But we can see that there is no genuine curvature contribution in four dimensions, because the term $\int_{cone} R$ always gives an area contribution. Our result is also consistent with their observation that in four dimensions the curvature contribution in Eqn-\ref{sw2} vanishes for a conformally coupled scalar for which $\xi'=0$. 

Eqn-\ref{sw0} and the above discussion are valid for any four dimensional field theory once we replace $\tilde M$ with $\Lambda_g$. 


\subsection{N=6}
In six dimensions there is genuine curvature contribution as predicted in \cite{Lewkowycz:2012qr}. We again get the area term contribution from the term,
\begin{equation}
(M^2 + \xi' R(S^2))^2 \ \int_{cone} R
\end{equation}
This is analyzed in the same way as in the four dimensional case and so we will not discuss it anymore. In six dimension there is one more contribution which can be written as,
\begin{equation}
(M^2 + \xi' R(S^2)) \ \int_{cone} (A R^2 + B R^2_{ab} +C R^2_{abcd})
\end{equation}
where $A$, $B$ and $C$ are pure dimensionless numbers. In our case we have to do these integrals on $C^2_{n}\times S^4$. Following \cite{Fursaev:1995ef} we get,
\begin{equation}
S_E \supset A (M^2 + \xi' R(S^4) ) \ ln\frac{\Lambda_{\tilde M}}{\Lambda} \int_{S^4} R(S^4)
\end{equation}
So the genuine curvature contributions start at six dimension. We can see how easy it is to write down the precise geometrical structures of the various universal terms in an unambiguous way. The above discussion goes through for any six dimensional field theory once we replace $\Lambda_{\tilde M}$ with the RG invariant scale $\Lambda_g$. Higher curvature terms do not arise for the same reason as in the case of four dimensions.

So we can see that these types of terms arise for any even dimensional field theory. There are only a finite number of terms in a given dimension. This is precisely the pattern observed in perturbative and holographic computations \cite{Hung:2011ta, Lewkowycz:2012qr}. Our method provides a field theoretic explanation of these facts and extends it to any even dimensional field theory. 

\section{Entanglement Entropy In Theories Interpolating Between Scale But Non-Conformally Invariant Fixed Points}
The reader may have noticed that throughout the discussion we needed only a constant dilaton field because we need a constant rescaling of the cutoff. With a constant dilaton field the symmetry which is preserved along the RG flow is not the full Weyl group but only the subgroup of constant conformal transformations or scale transformations. So our method should be applicable also to theories which are described in the UV by scale invariant but non-conformally invariant theories. The only change is that the dilaton effective action will be different because the integrated trace of the stress tensor is different if a theory is only scale invariant but not conformally invariant. We saw that the universal part of the dilaton effective action is completely determined by the integrated trace of the stress tensor in the UV theory. So if the UV theory is only scale invariant but not conformally invariant then, for example in four dimensions, the dilaton effective action will have an extra term \cite{Dymarsky:2014zja, Nakayama:2011wq} proportional to,  $\tau e_{UV} \int_{cone}\sqrt h R^2(h)$, where $e$ is the coefficient of the $R^2$ term which appears in the integrated trace of the stress tensor in a scale invariant theory. This term is ruled out in a CFT by the Wess-Zumino consistency condition. So if we have a theory which is described in the UV by scale but non-conformally invariant fixed points then the entanglement entropy will have an extra term proportional to, 
\begin{equation}
n\frac{\partial}{\partial n}|_{n=1} \ e_{UV} \int_{cone} d^4x \sqrt h R^2(h) \ ln\frac{\Lambda_g}{\Lambda}
\end{equation}


Now in two and four dimensions it has been proved that every unitary scale invariant theory is conformally invariant \cite{Polchinski:1987dy, Luty:2012ww,Dymarsky:2013pqa, Dymarsky:2014zja, Bzowski:2014qja}. So we do not expect the above term to appear in the entanglement entropy of a four dimensional unitary field theory. The same statement has not been proved so far for higher dimensional field theories and so extra terms appearing in the scale anomaly could be important if we compute the entanglement entropy of, say, a six dimensional theory.

\section{Discussion}
Now let us summarize our main results. We have studied entanglement entropy in field theories which interpolate between a UV-CFT and a trivial IR-CFT. Renormalization group together with the trace anomaly matching gives powerful constraints on the entanglement entropy, which is completely non-perturbative. This method gives precise quantitative information about entanglement entropy in any field theory. At this stage we would like to point out a basic difference between the work of \cite{Komargodski:2011vj,Komargodski:2011xv} and what we have done in this paper. The proof of the $a$-theorem uses the unitarity of the field theory in an essential way. In our work, so far the unitarity has played no role. So our results are also valid for non-unitary field theories, if we assume that entanglement entropy makes sense in non-unitary theories. Another interesting thing will be to reproduce the results of \cite{Liu:2012eea} using this approach. 
 
 So far we have only talked about the even dimensional field theories. Our method loses most of its power when applied to an odd dimensional field theory, because there is no anomaly in three dimensions. The dilaton effective action in odd dimensions has been probed using holographic methods in \cite{Bhattacharyya:2012tc}. It will be interesting to see if one could say something useful about entanglement entropy form the dilaton effective action in odd dimension. Many things are known about entanglement entropy in odd dimensional field theories \cite{Jafferis:2010un, Jafferis:2011zi, Casini:2012ei, Liu:2012eea, Casini:2012ei, Klebanov:2012va, Klebanov:2011td,Agon:2013iva}. It will be very interesting if one could reproduce and extend those results in some simple way. Probably a field theoretic proof of $F$-theorem will open new ways of probing this. 
 
Another interesting direction to pursue is to calculate the effect of other anomalies on entanglement entropy, like gravitational anomaly. For example in two dimensional CFTs gravitational anomaly means that the left-moving and right-moving central charges are not the same and there a term in the entanglement entropy proportional to the difference between the two central charges \cite{Holzhey:1994we, Castro:2014tta}. It will be interesting to see how this changes for a general non-conformal field theory. We hope to return to this issue in the near future.\\\\

\acknowledgments 
It is a great pleasure to thank Jyotirmoy Bhattacharya, Justin David, Simeon Hellerman, Dileep Jatkar, Jonathan Maltz, Yu Nakayama, Djordje Radicevic, Ashoke Sen, Edgar Shaghoulian, Steve Shenker, Aninda Sinha and Tadashi Takayanagi for helpful discussion and correspondence. I would also like to thank especially Dileep Jatkar, Djordje Radicevic, Steve Shenker, Aninda Sinha and Tadashi Takayanagi for very helpful comments on the draft. I would also like to thank the string theory group at Harish-Chandra Research Institute, India where part of this work was done. This work was supported by World Premier International Research Center Initiative (WPI), MEXT, Japan.  

\section{Note Added}
In a previous version of the paper results were stated for non-trivial IR-CFTs also. Those results are not valid because of the infrared divergences which arise when the IR-CFT is non-trivial. It appears that one has to add an extra contribution which can be interpreted as the contribution to the entanglement entropy from the IR CFT. The justification of this using the current approach will be part of a forthcoming publication. I am grateful to John Cardy for explaining this to me in detail and Tadashi Takayanagi for discussion on this matter.




\end{document}